\documentclass{article} 
 
\usepackage{amsmath,amssymb,amsthm,latexsym} 
 
\usepackage{stmaryrd,wasysym,upgreek,mathrsfs,dsfont} 
\usepackage[english]{babel} 
\usepackage{graphicx,color} 
\newcommand{\R}{\mathbb{R}} 
\newcommand{\Ot}{\tilde\Omega} 
\usepackage[pdftex]{hyperref} 

\newtheorem{lemma}{Lemma}[section]
\newtheorem{theorem}{Theorem}[section]
\newcommand{\beq}{\begin{equation}}
\newcommand{\eeq}{\end{equation}}
\newcommand{\beqa}{\begin{eqnarray}}
\newcommand{\eeqa}{\end{eqnarray}}

\newcommand{\e}{\varepsilon}

\newcommand{\cI}{{\cal{I}}}
\newcommand{\cR}{{\cal{R}}}
\newcommand{\cG}{{\cal{G}}}

\begin{document} 
 \title{Non-Commutative Complete Mellin Representation for  Feynman Amplitudes} 
\author{R. Gurau$^{(1)}$, A.P.C. Malbouisson$^{(2)}$\\ 
V. Rivasseau$^{(1)}$,
  A. Tanas{\u a}$^{(1)}$\\ 
1) Laboratoire de Physique Th\'eorique\\
CNRS UMR 8627, b\^at.\ 210\\ 
Universit\'e Paris XI,  F-91405 Orsay Cedex, France\\ 
2) Centro Brasileiro de Pesquisas F\'{\i}sicas,\\
Rua Dr.~Xavier Sigaud, 150, \\ 
22290-180 Rio de Janeiro, RJ, Brazil}
 
\maketitle 
\begin{abstract} 
We extend the complete Mellin (CM) representation of Feynman amplitudes 
to the non-commutative quantum field theories. This representation is a versatile tool. 
It provides a quick proof of meromorphy of Feynman amplitudes in parameters 
such as the dimension of space-time. In particular it 
paves the road for the dimensional renormalization of these theories. 
This complete Mellin representation also allows the study of asymptotic behavior 
under rescaling of arbitrary subsets of external invariants of any Feynman amplitude. 
\end{abstract} 
 
\section{Introduction} 

Recently non-commutative quantum field theories such as the $\phi^{\star 4}$ and
the Gross-Neveu$_2$ models have  been shown renormalizable \cite{GrWu03-1,GrWu03-2,GrWu04-1,Rivasseau2005bh,xphi4-05,RenNCGN05}
provided the propagator is modified to obey Langmann-Szabo duality \cite{LaSz}. These theories
are called ``vulcanized", and in \cite{gurauhypersyman,RivTan} the Feynman-Schwinger parametric 
representation was extended to them. For recent reviews, see \cite{RivTour,Riv}. 

In this paper we perform a further step, generalizing to these vulcanized NCQFT the 
complete Mellin (CM) representation.  This CM representation is derived from the 
parametric representation established in \cite{gurauhypersyman,RivTan} and provides a starting point 
for the study of dimensional renormalization and of the asymptotic behavior of Feynman amplitudes
under arbitrary rescaling of external invariants.

The study of asymptotic behaviors in 
conventional (commutative) field theories started in the 1970's with the papers \cite{bergere3,
bergere1, bergere2} which use the BPHZ
renormalization scheme. 
Their approach is based on the Feynman-Schwinger parametric representation of
amplitudes. This representation involves ``Symanzik" or ``topological" polynomials
associated to the Feynman diagram. 
In commutative theories these ``topological polynomials" are written as sums 
over the spanning trees or 2-trees of the diagram \cite{nakanishi, itzykson}. 
The corresponding mathematical theory goes
back to the famous tree matrix theorem of Birkhoff (see \cite{Abdesselam} and references therein).

In vulcanized NCQFTs, propagators are no longer based on the heat kernel
but on the Mehler kernel. The kernel being still quadratic in position space, explicit integration 
over all space variables is still possible. It leads to the 
NCQFT parametric representation. This representation no longer 
involves ordinary polynomials in the Schwinger parameters, 
but hyperbolic polynomials \cite{gurauhypersyman}. As
diagrams in NCQFT are ribbon graphs, these hyperbolic polynomials contain 
richer topological information than in the commutative case. In particular
they depend on the genus of the Riemann surface on which the graphs are defined.

Returning to the Mellin transform technique, it was
introduced in commutative field theory to prove theorems on the asymptotic
expansion of Feynman diagrams. The general idea 
was to prove the existence of an asymptotic series in powers of $\lambda $ and
powers of logarithms of $\lambda $, under rescaling by $\lambda$ of 
some (Euclidean) external invariants associated to the Feynman
amplitudes. The Mellin transform with respect to
the scaling parameter allows to obtain this result in some cases and to 
compute the series coefficients, even for renormalized diagrams \cite{bergerelam}.
But it did not work for\textit{ arbitrary} subsets of invariants.

In the subsequent years, the subject advanced further. 
The rescaling of internal squared masses (in order to study the infrared behavior 
of the amplitudes) was treated in \cite{malbouisson7}. There the concept of ``FINE" 
polynomials was introduced (that is, those being factorizable in each Hepp sector \cite{hepp}
of the variables\footnote{A Hepp sector is a complete ordering of the Schwinger parameters.}). It was then argued that the Mellin transform may be ``desingularized",
which means that the integrand of the inverse Mellin transform (which gives back the  
Feynman amplitude as a function of $\lambda $) has a meromorphic structure, so that the
residues of its various poles generate the asymptotic expansion in 
$\lambda $. However, this is not the case under arbitrary rescaling, 
because in many diagrams the FINE property simply does not occur. 

A first solution
to this problem was presented in \cite{malbouisson7} by introducing the
so-called ``multiple Mellin" representation, which consists in
splitting the Symanzik polynomials in a certain number of pieces, each one
of which having the FINE property. Then, after scaling by the parameter
 $\lambda $, an asymptotic expansion can be obtained as a sum over all Hepp
sectors. This is always possible if one adopts, as done in 
\cite{malbouisson, rivasseau, malbouisson5}, the extreme point of view to
split the Symanzik polynomials in all its monomials. Moreover, this apparent
complication is compensated by the fact that one can dispense with the use of 
Hepp sectors altogether. This is called the
``complete Mellin" (CM) representation.

The CM representation provides a general proof of the existence of an
asymptotic expansion in powers and powers of logarithms of the scaling
parameter in the most general case. Moreover the integrations over the 
Schwinger parameters can be explicitly performed, and we are left with the pure
geometrical study of convex polyhedra in the Mellin variables. The results of \cite{malbouisson7} are
obtained in a simpler way \cite{malbouisson}, and asymptotic expansions are computed in a more
compact form, without any division of the integral into Hepp sectors.

Moreover the CM representation allows a unified
treatment of the asymptotic behavior of both ultraviolet convergent and
divergent renormalized  amplitudes. Indeed, as shown in  \cite{malbouisson,rivasseau}, 
the renormalization procedure does not alter the algebraic
structure of integrands in the CM representation. It only changes the set of
relevant integration domains in the Mellin variables.
The method allows the study  of
dimensional regularization \cite{rivasseau,malbouisson5} and of the infrared behavior
of amplitudes relevant to critical phenomena \cite{malbouisson6}.

The CM representation is up to now the only tool which provides 
asymptotic expansions of Feynman amplitudes in powers and powers of logarithms in the most general
rescaling regimes. More precisely consider a Feynman amplitude $G(s_{k})$ written in terms
of its external invariants $s_{k}$ (including particle masses), and an asymptotic regime defined by 
\begin{equation}
s_{k}  \rightarrow \lambda ^{a_k} {s}_{k},  
\label{scaling1}
\end{equation}
where $a_k$ may be positive, negative, or zero.
Let $\lambda $ go to infinity  (in this way both ultraviolet and infrared behaviors 
are treated on the same footing). 
Then $G(s_{k})$ as a function of $\lambda $ has an 
asymptotic expansion of the form: 
\begin{equation}
G(\lambda;s_{k} )=\sum_{p=p_{\text{max}}}^{-\infty }\sum_{q=0}^{q_{\text{max}%
}(p)}G_{pq}(s_k)\lambda ^p\ln ^q\lambda ,
\label{expansao3}
\end{equation}
where $p$  runs over decreasing rational values, with $p_{max}$ as leading power, and $q$, for 
a given $p$, runs over a finite set of non-negative integer values. 

The CM representation proves this result for
any commutative field theory, including gauge theories in arbitrary gauges \cite{linhares}.

The above general result should hold {\it mutatis mutandis} for non-commutative 
field theories. In this paper we provide a starting point for this analysis by constructing 
the CM representation for $\phi^{*4}$ (Theorem \ref{cmrep} below). Moreover using this representation 
we check meromorphy of the Feynman amplitudes in the dimension of space-time (Theorem \ref{meromorph}). 

The main difference with the commutative case is that this integral representation (previously true 
in the sense of \textit{functions} of the external invariants)
now holds only in the sense of {\it distributions}.  Indeed
the distributional character of commutative amplitudes (in momentum space) 
reduces to a single overall $\delta$-function of momentum conservation. This is no longer true
for vulcanized NCQFT amplitudes, which must be interpreted as distributions to be smeared
against test functions of the external variables.

\section{Commutative CM representation} 
 
For simplicity we consider a scalar Feynman amplitude.
To get the CM representation \cite{malbouisson}, rewrite the Symanzik polynomials as  
\begin{equation} 
U(\alpha )=\sum_j\prod_{\ell=1}^L\alpha_\ell^{u_{\ell j}}\equiv \sum_jU_j,\;\;\qquad 
V(\alpha )=\sum_k s_k\left( \prod_{\ell=1}^L\alpha _\ell^{v_{\ell k}}\right) \equiv 
\sum_k V_k ,  \label{sym} 
\end{equation} 
where $j$ runs over the set of spanning trees and $k$ over the set of the 2-trees, 
\begin{equation} 
u_{\ell j}=\left\{  
\begin{array}{ll} 
0 & \text{if the line }\ell\text{ belongs to the 1-tree}\  j \\  
1 & \text{otherwise} 
\end{array} 
\right. 
\end{equation} 
and  
\begin{equation} 
v_{\ell k}=\left\{  
\begin{array}{ll} 
0 & \text{if the line }\ell\text{ belongs to the 2-tree}\  k\\  
1 & \text{otherwise.} 
\end{array} 
\right. 
\end{equation} 
 
The Mellin transform relies on the fact that for any function $f(u)$, piecewise smooth for $u>0$, if the integral  
\begin{equation} 
g(x)=\int_0^\infty du\,u^{-x-1}f(u) 
\end{equation} 
is absolutely convergent for $\alpha <$ Re $x<\beta $, then  for $\alpha <\sigma <\beta $ 
\begin{equation} 
f(u)=\frac 1{2\pi i}\int_{\sigma -i\infty }^{\sigma +i\infty }dx\,g(x)\,u^x .
\end{equation} 

Consider $D$ the space-time dimension to be for the moment real positive. 
Taking  $f(u) = e^{-u}$, and applying to $u=V_k/U$ one gets  
\begin{equation} 
e^{-V_k/U}=\int_{\tau _k}\Gamma (-y_k)\left( \frac{V_k}U\right) ^{y_k}, 
\label{gamayk} 
\end{equation} 
where $\int_{\tau _k}$ is a short notation for $\int_{-\infty }^{+\infty }%
\frac{d(\text{Im }y_k)}{2\pi }$, with Re $y_k$ fixed at $\tau _k<0$. We may 
now recall the identity 
\begin{equation} 
\Gamma (u)\left( A+B\right) ^{-u}=\int_{-\infty }^\infty \frac{d(\text{Im }x)%
}{2\pi }\Gamma (-x)A^x\Gamma (x+u)B^{-x-u} . \label{gamauamaisb} 
\end{equation} 
Taking $A\equiv U_1(x)$ and $B\equiv U_2+U_3+\cdots $ and using 
iteratively the identity above, leads, for $u=\sum_k y_k+D/2$, to 
\begin{equation} 
\Gamma \left( \sum_k y_k+\frac D2\right) U^{-\sum_k y_k-\frac D2}=\int_\sigma 
\prod_j\Gamma (-x_j)U_j^{x_j},  \label{gamayk2} 
\end{equation} 
with Re $x_j=\sigma _j<0$, Re $\left( \sum_k y_k +\frac D2\right) =\sum_k\tau 
_k+\frac D2>0$, and $\int_\sigma $ means $\int_{-\infty }^{+\infty }\prod_j
\frac{d(\text{Im }x_j)}{2\pi }$ with $\sum_jx_j+\sum_k y_k =-\frac D2$. Then, 
using (\ref{gamayk}) and (\ref{gamayk2}), the amplitude is written as  
\begin{equation} 
I_\cG (s_k ,m_l^2)=\int_\Delta \frac{\prod_j\Gamma (-x_j)}{\Gamma (-\sum_jx_j)}
\prod_k s_k ^{y_k }\Gamma (-y_k)\int_0^\infty \prod_l\,d\alpha _l\,\alpha 
_l^{\phi _l-1}\,e^{-\sum_l\alpha _lm_l^2},  \label{equacao} 
\end{equation} 
where  
\begin{equation} 
\phi _i\equiv \sum_ju_{ij}x_j+\sum_k v_{i k }y_k+1 . 
\end{equation} 
The symbol $\int_\Delta $ means integration over the independent 
variables $\frac{\text{Im }x_j}{2\pi }$, $\frac{\text{Im }y_k }{2\pi }$
in the convex domain $\Delta$ defined by ($\sigma $ and $\tau $ standing 
respectively for $\text{Re }x_j$ and $\text{Re }y_k$):  
\begin{equation} 
\Delta =\left\{ \sigma ,\tau \left|  
\begin{array}{l} 
\sigma _j<0;\;\tau _k<0;\;\sum_jx_j+\sum_k y_k =-\frac D2; \\  
\forall i,\;\text{Re }\phi _i\equiv \sum_ju_{ij}\sigma _j+\sum_k v_{ik }\tau 
_k +1>0 
\end{array} 
\right. \right\} \ . \label{domain} 
\end{equation} 
This domain $\Delta$ is non-empty as long as $D$ is positive and 
small enough so that every subgraph of $G$ has convergent power counting \cite{malbouisson},
hence in particular for the $\phi^4$ theory it is always non empty for any graph
for $0<D<2$. 
 
The $\alpha $ integrations may be performed, using the well-known 
representation for the gamma function, so that we have  
\begin{equation} 
\int_0^\infty d\alpha _l\,e^{-\alpha _lm_l^2}\alpha _l^{\phi _l-1}=\Gamma 
\left( \phi _l\right) \left( m_l^2\right) ^{-\phi _l}  \label{alphaint} 
\end{equation} 
and we finally get the CM representation of the amplitude in 
the scalar case:  
\begin{equation} 
I_G(s_k ,m_l^2)=\int_\Delta \frac{\prod_j\Gamma (-x_j)}{\Gamma (-\sum_jx_j)}%
\prod_k s_k ^{y_k }\Gamma (-y_k )\prod_l\left( m_l^2\right) ^{-\phi _l}\Gamma 
\left( \phi _l\right) .  \label{Mellin1} 
\end{equation} 
This representation can now be extended to complex values of $D$. For instance for a massive
$\phi^4$ graph, it is analytic in $D$ for $\Re D < 2$, and meromorphic in $D$ in the whole complex plane 
with singularities at rational values; furthermore its dimensional analytic continuation has
the same unchanged $CM$ integrand but translated integration contours \cite{rivasseau,malbouisson5}.

\section{Non-commutative parametric representation} 
\label{hyperbo} 
 
Let us summarize the results of \cite{gurauhypersyman}. 
Define the antisymmetric matrix $\sigma$ as 
\begin{align} 
\sigma=\begin{pmatrix} \sigma_2 & 0 \\ 0 & \sigma_2 \end{pmatrix} \mbox{ with}
\ \sigma_2=\begin{pmatrix} 0 & -i \\ i & 0 \end{pmatrix}\ . 
\end{align} 
The $\delta-$functions appearing in the vertices can be 
rewritten as an integral over some new variables $p_V$ called {\it hypermomenta},
{\it via} the relation 
\begin{align} 
\label{pbar1} 
\delta(x_1 -x_2+x_3-x_4 ) = \int  \frac{d p'}{(2 \pi)^4} 
e^{ip' (x_1 -x_2 +x_3 -x_4)}
=\int  \frac{d p}{(2 \pi)^4} 
e^{p \sigma (x_1-x_2+x_3-x_4)}. 
\end{align}

Let $\cG$ be a ribbon graph. Choosing a particular root vertex $\bar{V}$,
the parametric representation for the amplitude of $\cG$ is
expressed in terms of $t_\ell = \tanh \alpha_\ell /2 $ (where $\alpha_\ell$ is the former
Schwinger parameter) as
\begin{align} 
\label{a-condens} 
{\cal A}_{\cG} =& \int_0^1 \prod_{\ell} d t_\ell (1-t_{\ell}^2)^{\frac D2 -1}  
\int d x d p e^{-\frac {\Omega}{2} X G X^t} 
\end{align} 
where $D$ is the space-time dimension, $X$ summarizes all positions and hypermomenta 
and $G$ is a certain quadratic form. Calling
$x_e$ and $p_{\bar{V}}$ the external variables and $x_i, p_i$ 
the internal ones, we decompose $G$ into an internal  
quadratic form $Q$, an external one $M$ and a coupling part $P$ so that  
\begin{align} 
\label{defX} 
X =& \begin{pmatrix} 
x_e & p_{\bar{V}} & x_i & p_i\\ 
\end{pmatrix} , \ \  G= \begin{pmatrix} M & P \\ P^{t} & Q \\ 
\end{pmatrix}\ , 
\end{align} 
Performing the Gaussian integration over all internal variables one gets the non-commutative
parametric representation: 
\begin{align} 
\label{aQ} 
{\cal A}_{\cG}  =&  \int \prod_{\ell}d t_\ell (1-t_{\ell}^2)^{\frac D2 -1} 
\frac{1}{\sqrt{\det Q}} 
e^{-\frac{\Ot}{2} 
\begin{pmatrix} x_e & p_{\bar V} \\ 
\end{pmatrix} \big[ M-P Q^{-1}P^{t}\big ] 
\begin{pmatrix} x_e \\  p_{\bar V}\\ 
\end{pmatrix}}\ . 
\end{align} 
This representation leads to new polynomials $HU_{\cG, \bar{V}}$ and $HV_{\cG, \bar{V}}$
 in the $t_\ell$ ($\ell =1,\ldots, L$) variables, analogs  
of the Symanzik polynomials $U$ and $V$ of the commutative case,
through
\begin{align} 
\label{polv} 
{\cal A}_{{\cG}}  (\{x_e\},\;  p_{\bar V}) =& K'  \int_{0}^{1} 
\prod_{\ell}d t_\ell (1-t_{\ell}^2)^{\frac D2 -1} 
\frac{e^{-  \frac {HV_{\cG, \bar{V}} ( t , x_e , p_{\bar v})}{HU_{\cG, \bar{V}} ( t )}}}
{HU_{\cG, \bar{V}} ( t )^{\frac D2}} .
\end{align} 
 
The main results of \cite{gurauhypersyman,RivTan} are 
 
\begin{itemize} 
 \item The polynomial $HU_{\cG, \bar{V}}$ and the real part of $HV_{\cG, \bar{V}}$  
have a \emph{positivity property}. They are sums of monomials 
with positive integer coefficients, which are squares of Pfaffians  
with integer entries.
 
\item {Leading terms} can be identified in a given ``Hepp sector'',  
at least for \textit{orientable graphs}. 
In $HU_{\cG, \bar{V}}$ they correspond to hyper-trees 
which are the disjoint union of a tree in the direct graph and an other tree in the dual graph. Any 
connected graph has such hypertrees. Similarly ``hyper-two-trees'' 
govern the leading behavior of $HV_{\cG, \bar{V}}$ in any Hepp sector. 
\end{itemize} 

Let us relabel the $2L$ internal positions of $\cG$ as $L$ short and
$L$ long variables in the terminology of \cite{xphi4-05}.
It has been shown in \cite{gurauhypersyman} that  for the Grosse-Wulkenhaar $\phi^{\star 4}$ model:
\beqa\label{hypnoncanpfaff}
HU_{\cG,{\bar V}} (t) &=&   \sum_{{K_U}= I\cup J, \  n + |{K_U}|\; {\rm odd}}  s^{2g-k_{{K_U}}} \ n_{{K_U}}^2
\prod_{\ell \not\in I} t_\ell \prod_{\ell' \in J} t_{\ell'}\ 
\nonumber \\
&=&\sum_{{K_U}}a_{K_U} \prod_\ell t_\ell^{u_{\ell {K_U}}}\equiv 
\sum_{K_U} HU_{K_U} .
\eeqa
where  
\begin{itemize}

\item $I$ is a subset of the first $L$ indices (corresponding to the short variables) with $\vert I \vert $ elements,
and $J$ a subset of the next $L$ indices (corresponding to the long ones) with $\vert J\vert $ elements.
 
\item $B$ is the antisymmetric part of the quadratic form $Q$ restricted to these $2L$ short and long
variables (i.e. omitting hypermomenta)

\item $n_{{K_U}}=\mathrm{Pf}(B_{\widehat{{K_U}}})$ is the Pfaffian of the antisymmetric matrix 
obtained from $B$ by deleting  the lines and columns in the set ${K_U}= I \cup J$.

\item $k_{{K_U}}$ is $ \vert {K_U}\vert - L - F +1$, where $F$ is the number
of faces of the graph, $g$ is the genus of $\cG$ and $s$ is a constant.

\item $ a_{K_U} = s^{2g-k_{{K_U}}}n_{{K_U}}^2$, and
\begin{equation}\label{ul0}
u_{\ell {K_U}}=\left\{  
\begin{array}{ll} 
0 & \mbox{ if } \ell \in I \mbox{ and } \ell\notin J \\  
1 & \mbox{ if } (\ell \notin I \mbox{ and } \ell\notin J) \mbox{ or } (\ell \in I \mbox{ and } \ell\notin J)\\
2 &  \mbox{ if } \ell \notin I \mbox{ and } \ell\in J
\end{array} 
\right. .
\end{equation}

\end{itemize}

 The second polynomial $HV$ has both a real part $HV^\cR$ and an imaginary part
 $HV^\cI$, more difficult to write down. We need to introduce beyond $I$ and $J$
as above a particular line $\tau \notin I$ which is the analog of a two-tree cut. 
We define $\mathrm{Pf}(B_{\hat{{K_V}}\hat{\tau}})$ as the Pfaffian of the matrix obtained from $B$ by deleting 
the lines and columns in the sets $I$, $J$ and $\tau$. Moreover we
define $\epsilon_{I,\tau}$ to be the signature of the permutation obtained
from $(1,\ldots, d)$ by extracting the positions belonging to $I$ and
replacing them at the end in the order
\beqa
1,\dotsc ,d\rightarrow 1,\dotsc,\hat{i_1},\dotsc ,\hat{i_{\vert I\vert }},\dotsc
,\hat{i_{\tau}},\dotsc ,d, i_{\tau},i_{\vert I\vert }\dotsc, i_1\,.
\eeqa
where $d$ is the dimension of the matrix $Q$. Then:
\beqa
HV^\cR_{\cG,{\bar V}}&=&
 \sum_{{K_V}= I \cup J }\prod_{\ell \notin I} t_\ell
 \prod_{\ell' \in J} t_{\ell'}
\Big{[}\sum_{e_1}x_{e_1}\sum_{\tau\notin {K_V}}P_{e_1\tau}\epsilon_{{K_V}\tau}
\mathrm{Pf}(B_{\hat{{K_V}}\hat{\tau}})\Big{]}^2 \, .
\label{HVgv1}\nonumber\\
&=&
\sum_{K_V} s^\cR_{K_V}\left( \prod_{\ell=1}^L t_\ell^{v_{\ell {K_V}}}\right) \equiv 
\sum_{K_V} HV_{K_V}^\cR
\eeqa
where
\beqa s_{K_V}^\cR= \left(
\sum_e x_e \sum_{\tau\notin {K_V}} P_{e\tau} \e_{{K_V} \tau} 
\mathrm{Pf}(B_{\hat{{K_V}}\hat{\tau}})
\right)^2 \eeqa
and $v_{\ell {K_V}}$ given by the same formula as $u_{\ell K_U}$.

The imaginary part involves {\it pairs} of lines $\tau, \tau'$ and corresponding
signatures (see \cite{gurauhypersyman,RivTan} for details):
\beqa \label{imaghv}
HV_{\cG,\bar V}^\cI&=&\sum_{{K_V}= I \cup J }\prod_{\ell \notin I} t_\ell
 \prod_{\ell' \in J} t_{\ell'}\nonumber \\
&&
\epsilon_{K_V}\mathrm{Pf}(B_{\hat{{K_V}}})
\Big{[}\sum_{e_1,e_2} \Big{(}
\sum_{\tau\tau'}P_{e_1\tau}\epsilon_{{K_V}\tau\tau'}\mathrm{Pf}(B_{\hat{{K_V}}\hat{\tau}\hat{\tau'}})
P_{e_2\tau'}\Big{)} x_{e_1}\sigma x_{e_2}\Big{]}\,.\nonumber
\\
&=&
\sum_{K_V} s^\cI_{K_V}\left( \prod_{\ell=1}^L t_\ell^{v_{\ell {K_V}}}\right) \equiv 
\sum_{K_V} HV_{K_V}^\cI
\eeqa
where
\beqa s_{K_V}^\cI=\epsilon_{K_V} \mathrm{Pf}(B_{\hat{{K_V}}})
\left(
\sum_{e, e'} (\sum_{\tau, \tau'}   P_{e \tau} \e_{{K_V} \tau \tau'} 
\mathrm{Pf}(B_{\hat{{K_V}}\hat{\tau}\hat{\tau'}}) P_{e'\tau'})x_e \sigma x_{e'}
\right) . \eeqa

\medskip
A similar development exists for the LSZ model (\cite{RivTan}), although more involved. 
The first polynomial in this case writes 
\beqa 
\label{pol-f} 
HU_{\cG, {\bar V}} (t) &=&   
\sum_{\substack{I\subset \{1\dotsc L\},\\ n + |I|\; {\rm odd}}}  s^{2g-k_{I}} \ n_{I}^2 
\prod_{l \in I} \frac{1+t_\ell^2}{2t_\ell} \prod_{l' \in \{1,\ldots,L\}} t_{l'}
\nonumber \\
&=&\sum_{{K_U}= I \cup J}a_{K_U} \prod_\ell t_\ell^{u_{\ell {K_U}}}\equiv 
\sum_{K_U} HU_{K_U} .
\eeqa 
Again $I$ runs over the subsets of $\{ 1, \ldots, L\}$. But $J$ now runs over subsets of 
$I$ (representing the expansion
of $\prod_{\ell \in I} [1 + t_\ell^2]$), and one has again
\begin{equation} 
\label{ul}
u_{\ell {K_U}}=\left\{  
\begin{array}{ll} 
0 & \mbox{ if } \ell\in I \mbox{ and } \ell \in J\\  
1 & \mbox{ if } \ell\notin I\\
2 &  \mbox{ if } \ell\in I \mbox{ and } \ell \notin J  
\end{array} 
\right. \ \ , \ \  
a_{K_U} = \frac{1}{2^{|I|}}s^{2g-k_{I}}n_I^2.
\end{equation} 
There are similar expressions for the second polynomial.

To summarize, the main differences of the NC parametric representation 
with respect to the commutative case are:
\begin{itemize}
\item the presence of the constants $a_j$ in the form
(\ref{hypnoncanpfaff}) of $HU$,
\item the presence of an imaginary part $i\, HV^\cI$ in
$HV$,  see (\ref{imaghv}),
\item the fact that the parameters $u_{\ell j}$ and $v_{\ell k}$ in the formulas above
can have also the value $2$ (and not only $0$ or $1$).
\end{itemize}

\section{Non-commutative CM representation}

For the real part $HV^\cR$ of $HV$ one uses again the identity \eqref{gamayk} as
\begin{equation} 
\label{yR}
e^{-HV^\cR_{K_V}/U}=\int_{\tau_{K_V}^\cR}\Gamma (-y_{K_V}^\cR)\left( \frac{HV^\cR_{K_V}}U\right) ^{y_{K_V}^\cR}, 
\end{equation} 
which introduces the set of Mellin parameters $y_{K_V}^\cR$.

However for the imaginary part one cannot apply anymore the same identity. Nevertheless it remains
true {\bf in the sense of
distributions}. More precisely we have for $HV^\cI_{K_V}/U >0$ and $-1 < \tau_{K_V}^\cI <0$ 
\begin{equation} 
\label{yI}
e^{-i\, HV^\cI_{K_V}/U}=\int_{\tau_{K_V}^\cI}\Gamma (-y_{K_V}^\cI)\left( \frac{i\, HV^\cI_{K_V}}U\right) ^{y_{K_V}^\cI}.
\end{equation} 
The proof is given in Appendix B.

Note that this introduces another  set of Mellin parameters $y_{K_V}^\cI$. 
The distributional sense of formula \eqref{yI} is
a translation of the distributional form of a vertex contribution,
and the major difference with respect to the commutative case.

\medskip

For the polynomial $HU$ one can use again the formula \eqref{gamayk2} which rewrites here as
\begin{equation} 
\Gamma \left( \sum_{K_V}y_{K_V}+\frac D2\right) (HU)^{-\sum_{K_V} (y_{K_V}^\cR+ y_{K_V}^\cI) -\frac D2}=\int_\sigma 
\prod_{K_U}\Gamma (-x_{K_U})U_{K_U}^{x_{K_U}},  \label{x} 
\end{equation}
Note that the r\^ole of the Mellin parameters  $y_{K_V}$ of the commutative case is
now played by the sum $y_{K_V}^\cR + y_{K_V}^\cI$. 

\medskip

As in the commutative case, we now insert the distribution formulas
\eqref{yR}, \eqref{yI} and \eqref{x} in the general form of the Feynman
amplitude. This gives
\beqa 
\label{aproape}
{\cal A}_G &=&{\rm K'}\int_\Delta \frac{\prod_{K_U} a_{K_U}^{x_{K_U}} \Gamma (-x_{K_U})}{\Gamma (-\sum_{K_U}x_{K_U})}
\left( \prod_{K_V} (s_{K_V}^\cR)^{y_{K_V}^\cR}\Gamma (-y_{K_V}^\cR) \right)
\nonumber\\
&&\left( \prod_{K_V} (s_{K_V}^\cI)^{y_{K_V}^\cI}\Gamma (-y_{K_V}^\cI) \right)\int_0^1 \prod_{\ell =1}^L dt_\ell (1-t_\ell^2)^{\frac D2 -1}
t_\ell^{\phi_\ell -1}
\eeqa
where  
\begin{equation} 
\phi_\ell\equiv \sum_{K_U} u_{\ell {K_U}}x_{K_U}+\sum_{K_V}
(v^\cR_{\ell {K_V}}y_{K_V}^\cR+v^\cI_{\ell {K_V}}y_{K_V}^\cI) +1 .
\end{equation} 
Here $\int_\Delta $ means integration over the
variables $\frac{\text{Im }x_{K_U}}{2\pi i}$, $\frac{\text{Im }y_{K_V}^\cR}{2\pi i}$ and 
$\frac{\text{Im }y_{K_V}^\cI}{2\pi i}$, where
$\Delta $ is the convex domain:  
\begin{equation} 
\Delta =\left\{ \sigma ,\tau^\cR, \tau^\cI \left|  
\begin{array}{l} 
\sigma _{K_U}<0;\;\tau^\cR_{K_V}<0;\; -1 < \tau^\cI_{K_V}<0;\;
\\
\sum_{K_U}x_{K_U}+\sum_{K_V} (y_{K_V}^\cR+ y_{K_V}^\cI)=-\frac D2; \\  
\forall \ell,\;\text{Re }\phi_\ell\equiv \sum_{K_U}u_{\ell {K_U}}\sigma _{K_U}\\
+\sum_{K_V}
(v^\cR_{\ell {K_V}}\tau^\cR_{K_V}+v^\cI_{\ell {K_V}}\tau^\cI_{K_V})+1>0 
\end{array} 
\right. \right\}  \label{nc-domain} 
\end{equation} 
and $\sigma $, $\tau^\cR $ and $\tau^\cI $ stand 
for $\text{Re }x_{K_U}$, $\text{Re }y_{K_V}^\cR$   and $\text{Re }y_{K_V}^\cI$.

The $dt_\ell$ integrations in \eqref{aproape} may be performed using the  
representation for the beta function
\beqa
\label{beta}
\int_0^1 \prod_{\ell =1}^L dt_\ell (1-t_\ell^2)^{\frac D2 -1}
t_\ell^{\phi_\ell -1}=\frac 12 \beta (\frac{\phi_\ell}2, \frac D2).
\eeqa
Furthermore one has
$$  \beta (\frac{\phi_\ell}2, \frac D2) =\frac {\Gamma (\frac{\phi_\ell}2)
  \Gamma (\frac D2)}{\Gamma (\frac{\phi_\ell+D}2)}. $$
This representation is convergent for  $0<\Re D<2$ and we get:
  
\begin{theorem}\label{cmrep}
Any Feynman amplitude of a $\phi^{\star 4}$ graph
is analytic at least in the strip $0<\Re D<2$ where it
admits the following CM representation 
\beqa 
\label{final}
{\cal A}_\cG &=&{\rm K'}\int_\Delta \frac{\prod_{K_U} a_{K_U}^{x_{K_U}} \Gamma (-x_{K_U})}{\Gamma (-\sum_{K_U}x_{K_U})}
\left( \prod_{K_V} (s_{K_V}^\cR)^{y_{K_V}^\cR}\Gamma (-y_{K_V}^\cR) \right)
\nonumber\\
&& \left( \prod_{K_V} (s_{K_V}^\cI)^{y_{K_V}^\cI}\Gamma (-y_{K_V}^\cI) \right)
\left( \prod_{\ell=1}^L \frac {\Gamma (\frac{\phi_\ell}2)
  \Gamma (\frac D2)}{2\Gamma (\frac{\phi_\ell+D}2)} \right).\nonumber\\
\eeqa
which holds as tempered distribution of the external invariants.
\end{theorem}
\noindent{\bf Proof}
We have to check that for any $\phi^{\star 4}$ graph, the domain $\Delta$ 
is non empty for $0<\Re D<2$. It is obvious to check that $\Delta$ is non empty for $0<\Re D<1$;
indeed since the integers $u$ and $v$ are bounded by 2, we can put an arbitrary single $x_{K_U}$
close to $-D/2$ and the others with negative real parts close to 0 and the conditions 
$\phi_\ell >0$ will be satisfied. With a little extra care, one also gets easily that
$\Delta$ is non empty for $0<\Re D<2$. Indeed one can put again one single $x_{K_U}$
close to $-D/2$, provided for any $\ell$ we have $u_{\ell,{K_U}} \ne 2$. For the ordinary 
Grosse-Wulkenhaar model one can always find ${K_U}={I\cup J}$
where $I=\{1,...,L\}$ is the full set of lines \cite{gurauhypersyman}, hence  $u_{\ell,{K_U}} \ne 2$ for any $\ell$
by (\ref{ul0}).
In the ``covariant" case treated in \cite{RivTan}, one can simply take ${K_U}$ with $J=I$.
Considering (\ref{ul}) ensures again obviously that  $u_{\ell,{K_U}} \ne 2$ for any $\ell$.

The analyticity statement is part of a more 
precise meromorphy statement proved in the next section.
\qed

\medskip
We have thus obtained the {\bf complete Mellin representation of Feynman
amplitudes for non-commutative QFT}. Let us
point out first that the distributional aspect does not affect
the last product, the one involving the dimensionality of space-time
$D$. Therefore, one can still prove the meromorphy of the Feynman amplitude
${\cal A}_\cG (D) $ in the space-time dimension $D$ 
essentially as in the commutative case. This is done in the next section.

Another difference is that the $dt_\ell$ integration brought $\beta$ 
functions (see \eqref{beta}), hence more complicated ratios of $\Gamma$ 
functions than in the commutative case. In particular 
singularities appear for negative $D$ which come from the integration near
$t_\ell =1$, hence the ``infrared side". They have absolutely no analog
in the commutative case and are due to the hyperbolic nature
of the Mehler kernel. At zero mass these singularities
occur for $D$ real negative even integer and in the massive case
they occur for $D$ real but sufficiently negative depending on the mass. 

\section{Meromorphy in the space-time dimension $D$} 
 
The complete Mellin representation \eqref{final} for non-commutative
theories  allows a quick proof of the meromorphy of the Feynman
amplitude ${\cal A}_G $ in the variable $D$, since $D$ appears only as argument
of $\Gamma$ functions which are meromorphic. 

\begin{theorem}\label{meromorph}
Any amplitude ${\cal A}_\cG $ is a tempered meromorphic distribution in $D$, that is 
 ${\cal A}_\cG $ smeared against any fixed Schwarz-class test function
of the external invariants yields a meromorphic function in $D$
in the entire complex plane, with singularities located among a discrete rational 
set $S_G$ which depends only on the graph $G$ not on the test function. 
In the $\phi^4$ case, no such singularity can occur in the strip $0<\Re D<2$, 
a region which is therefore a germ of analyticity common to the whole theory.
\end{theorem}

\noindent{\bf Proof}
 $\Gamma$ functions and their inverse
are meromorphic in the entire complex plane.
Formula (\ref{final}) is a tempered distribution 
which is the Fourier transform of a slow-growth function, see Appendix A.
It must be smeared against a Schwarz-class test function and by definition the result  
can be computed in Fourier space, as an ordinary integral of the product of
the slow-growth function by the rapid decaying Fourier transform of the test function.
At this stage the $D$-dependence occurs in fact in the integrand of an ordinary integral.

We can now copy the analysis of \cite{rivasseau,malbouisson5}. We introduce
general notations $z_\nu$ for the variables $x$ and $y$ and the lattices of polar varieties 
$\phi_{r}= -n_r$ where the different $\Gamma$ functions have their singularities. 
This defines convex cells, and singularities in $D$ can occur only when the hyperplane
$P(D)= \{\sum_\nu z_\nu = -D/2\}$ crosses the vertices of that lattice.
Because the coefficients $u$ and $v$ can only take integer values (in fact 0, 1 and 2)
this give a discrete set of rational values $S_G$ which depend on the graph, but not on the test-function. 
Out of these values,
to check that this integral is holomorphic around any dimension which does not cross the lattice vertices
is a simple consequence of commuting the Cauchy-Riemann $\partial_{\overline{D}}$ 
operator with the integral through Lebesgue's dominated convergence theorem.

Finally to check meromorphy, fix a particular $D_0 \in S_G$. We can write the Laurent series expansion of the 
integrand near that $D_0$ which starts as $a_{n_0} (D-D_0)^{-n_0}$. Multiplying
by $(D-D_0)^{n_0}$ we get a finite limit as $D \to D_0$, hence the isolated singularity at $D_0$
cannot be essential. Repeating the argument at each point of $S_G$ we obtain
meromorphy in the space time dimension $D$ in the whole complex plane $\mathbb{C}$,
apart from some discrete rational set where poles of finite order can occur.
\qed

Recall that this theorem is the starting point for the 
dimensional renormalization of NCQFTs 
which will be published elsewhere \cite{GurTan}.

\medskip
\noindent{\bf Appendix A: Proof of the formula \eqref{yI}}
\medskip

Let us return to formula (\ref{yI}). It follows from 
\begin{lemma}
For $-1 < s <0$, as tempered distribution in $u$
\begin{equation} 
\label{yIsimp}
e^{-iu}\chi (u)=\chi(u)
\int_{-\infty}^{+ \infty}\Gamma (-s-it)\left( iu \right) ^{s+it} \frac{d t}{2\pi }.
\end{equation}  
where $\chi$ is the Heaviside step function ensuring $u >0$.
\end{lemma}

\noindent{\bf Proof}
The left hand side is a locally integrable function bounded 
by one (together with all its derivatives) times the Heaviside function, so it is a tempered 
distribution. The right hand side is in fact a Fourier transform in $t$ but taken at a value $\log u$. 
To check that this is the same tempered distribution as $e^{-iu}\chi (u)$, let us apply 
it to a Schwartz test function $f$. Writing the usual representation for
the $\Gamma$ function we have to compute
\begin{equation} 
\label{ysimp2} I(f)=\frac{1}{2\pi }
\int_{0}^{\infty}  f(u) du 
 \int_{-\infty}^{+\infty} (iu)^{s+it} dt \int_{0}^{\infty} x^{-s-it-1} e^{-x}dx.
\end{equation} 
But the integrand is analytic in $x$ in the open upper right quarter of the complex plane. Let us
rotate the contour from $x$ to $ix$. The contribution of the quarter circle with radius sent
to infinity tends to zero because of the condition $-1 < s <0$, and we find 
\begin{eqnarray} 
\label{ysimp3} I(f)&=&\frac{1}{2\pi }
\int_{0}^{\infty}  f(u) du 
\int_{-\infty}^{+\infty} (iu)^{s+it} dt \lim_{R \to \infty}
\int_{0}^{R} (ix)^{-s-it-1} e^{-ix} i dx \nonumber
\\&=&\frac{1}{2\pi } \int_{-\infty}^{+\infty} dt  
\int_{-\infty}^{\infty}  f(e^v) e^{v(s+1)}  e^{itv} dv   \lim_{R\to \infty}\int_{0}^{R} x^{-s-it-1}e^{-ix} dx
\nonumber
\\&=&\frac{1}{2\pi } \int_{-\infty}^{+\infty} dt  \hat G (t)  \lim_{R\to \infty}\int_{0}^{R} 
  x^{-s-it-1}e^{-ix} dx
  \nonumber
\\&=& \lim_{R\to \infty}\int_{0}^{R}dx  e^{-ix}   x^{-s-1}  
\frac{1}{2\pi }\int_{-\infty}^{+\infty} dt  \hat G (t) e^{-it\log x}
\nonumber   
\\&=& \int_{0}^{\infty} f(x)  e^{-ix} dx\ .
\end{eqnarray} 
where $G(v) = f(e^v) e^{v (s+1)}$ is still rapid decay and
in the second line we change variables from $u$ to $v=\log u$ and use the 
definition of Fourier transform for distributions.
\qed

\medskip
\noindent{\bf Appendix B: The massive case}
\medskip

In the massive case we have to modify equation \eqref{beta} to include the additional
$e^{- \alpha_\ell m^2}$ per line, but unfortunately 
we have $e^{- \alpha_\ell} = \frac{1-t_\ell}{1+t_\ell}$ so that 
\eqref{beta} becomes
\beqa
\label{beta1}
\int_0^1 \prod_{\ell =1}^L dt_\ell   (1-t_\ell^2)^{\frac D2 -1}  \biggl(\frac{1-t_\ell}{1+t_\ell}\biggr)^{m^2}
t_\ell^{\phi_\ell -1}
=\frac 12 \beta_{m} (\frac{\phi_\ell}2, \frac D2).
\eeqa
where the modified ``$\beta_m$"  function is no longer an explicit quotient of $\Gamma$
functions, and depends of the mass $m$. 

It is rather easy to check that the analyticity and meromorphy properties
of this modified beta function are similar to those of the ordinary beta function,
at least for $\Re D >0$. Indeed
\beqa
\beta_{m} (\frac{\phi_\ell}2, \frac D2)-\beta (\frac{\phi_\ell}2, \frac D2)
=2\int_0^1 \prod_{\ell =1}^L dt_\ell   (1-t_\ell^2)^{\frac D2 -1}  \Big{[}
\biggl(\frac{1-t_\ell}{1+t_\ell}\biggr)^{m^2}-1
 \Big{]}
t_\ell^{\phi_\ell -1}.
\eeqa

The integral above is convergent for $t_\ell$ close to $1$ if $\Re D>0$ and for $t_\ell$ close to $0$ if $\Re \phi_l>-1$ so that it is an analytic function of $D$ and $\phi_l$ in this domain, hence $\beta$ and $\beta_m$
have the same singularity structure.
 
Nevertheless this CM representation becomes less explicit and therefore less attractive in this massive case.
Remark however that for $\Re D>0$, masses are not essential to the analysis of 
vulcanized NC field theories  which have no infrared divergencies and only ``half-a-direction" for 
their renormalization group anyway. As remarked earlier, masses can only push further away
the infrared singularities at negative $D$ that come form the hyperbolic nature of the Mehler kernel.
But it is not clear whether such infrared singularities at negative non commutative dimensions
have any physical interpretation.

\end{document}